\documentclass[a4paper,11pt]{article}
\usepackage{setspace}
\usepackage{amsmath}
\usepackage{graphicx}
\usepackage[square,numbers,sort&compress]{natbib}
\title{Mechanical model for a collagen fibril pair in extra cellular matrix}
\author{Chan Y.$^{1}$, Cox G.M.$^{1}$, Haverkamp R.G.$^{2}$, Hill J.M.$^{1}$\\
\footnotesize{$^{1}$ Nanomechanics Group, School of Mathematics and Applied Statistics,}\\
\footnotesize{University of Wollongong, Wollongong, NSW, 2522, Australia}\\
\footnotesize{$^{2}$ Institute of Technology and Engineering, Massey University,}\\
\footnotesize{Palmerston North 5331, New Zealand}}

\begin{document}

\maketitle

\abstract{In this paper, we model the mechanics of a collagen pair
in the connective tissue extracellular matrix that exists in
abundance throughout animals, including the human body. This
connective tissue comprises repeated units of two main structures,
namely collagens as well as axial, parallel and regular anionic
glycosaminoglycan between collagens. The collagen fibril can be
modeled by Hooke's law whereas anionic glycosaminoglycan behaves
more like a rubber-band rod and as such can be better modeled by the
worm-like chain model. While both computer simulations and continuum
mechanics models have been investigated the behavior of this
connective tissue typically, authors either assume a simple form of
the molecular potential energy or entirely ignore the microscopic
structure of the connective tissue. Here, we apply basic physical
methodologies and simple applied mathematical modeling techniques to
describe the collagen pair quantitatively. We find that the growth
of fibrils is intimately related to the maximum length of the
anionic glycosaminoglycan and the relative displacement of two
adjacent fibrils, which in return is closely related to the
effectiveness of anionic glycosaminoglycan in transmitting forces
between fibrils. These reveal the importance of the anionic
glycosaminoglycan in maintaining the structural shape of the
connective tissue extracellular matrix and eventually the shape
modulus of human tissues. We also find that some macroscopic
properties, like the maximum molecular energy and the breaking
fraction of the collagen, are also related to the microscopic
characteristics of the anionic glycosaminoglycan.

\bigskip
\section{Introduction}
\label{sec:1}

Structural polymers help to maintain the human body shape from large
external tractions. Collagen occurs extensively in all animals and
is the defining structural polymer, existing in the connective
tissue extracellular matrix (CTs), for example, skin, cartilage and
bone. It acts to support life against vigorous daily activities. CTs
are bridged and bonded by anionic glycosaminoglycan GAGs, such as
parallel rows of decoran, which are the only molecules in CTs apart
from protein fibres that can be visualized by an electron microscope
\cite{Scott1991,Scott1988,Svoboda}. A pair of segments on
neighboring collagen fibrils, which are linked by GAGs chains, is
termed a shape module and must deform reversibly to preserve the
general structure of the organism against various external stresses
\cite{Haverkamp2005}. The axial tension is transmitted and opposed
by protein fibres while compression is resisted by water-soluble
polysaccharide GAGs, e.g. chondroitin sulphate. These GAGs transmit
forces from the local area of molecules to global fibrils by
converting compression into disseminated tensile stress
\cite{Scott}. The mechanics involving the collagen and the elastin
fibril is well understood, i.e. they behave mainly like a Hookean
spring in the low stress limit, but the elasticity of GAGs'
molecules is still under intensive investigation \cite{Burr}.

Numerous computational and continuum mechanical methodologies have
been utilized to study the mechanics of such CTs
\cite{Redaelli,Akkus,Fritsch,Hellmich,Jager,Ji,Kotha,Wang1}.
However, most tend to ignore the microscopic details of CTs or
assume a simple form of the molecular potential energy in order to
make their models tractable. In particular, for a continuum
mechanical approach, generally a homogeneous structure of CTs is
assumed. Since the CTs comprise of collagen pairs, here we adopt
basic physical concepts and simple mathematical techniques to
investigate the mechanical properties of a single collagen pair.
While utilizing a quadratic energy form for the collagen, we
incorporate the statistical nature of GAGs into our model by
utilizing the worm-like chain model, which has both theoretically
and experimentally been proved to be applicable to wide ranges of
bio-macromolecules including unstructured DNA, RNA, GAGs and
polysaccharide \cite{Haverkamp2005,Baumann,Rief1,Rief2,Oesterhelt}.
We note that our work may form a theoretical basis for
experimentalists for their work on collagen pairs.

GAGs can go through conformational transitions, which are defined by
the sudden elongation of bio-molecules due to the change in their
atomic allocations without an increase in external forces. GAGs,
e.g. pectinate, have two distinct chair structures, namely
$^{4}C_{1}$ and $^{1}C_{4}$ \cite{Barrows}, separated by an energy
barrier of approximately $11$ kcal/mol \cite{Joshi}. Moreover, there
exists a boat conformation $^{1,4}B$ with an energy level of
approximately $5-8$ kcal/mol above the most stable $^{4}C_{1}$ chair
energy \cite{Dowd}. In the low energy configuration, nature prefers
the $^{4}C_{1}$ chair state over other possible states because it
possesses the minimum energy configuration in comparison to other
possible states. However, under an external stress, these
polysaccharides can undergo two conformational transitions
reversibly beyond some critical stresses \cite{Marszalek}. For
example, amylose undergoes its first conformational transition when
the applied force is around $200$ pN and its second transition when
the applied force is around $500$ pN. However, some polysaccharides
like pigskin DS only go through one transition, while other
polysaccharides like poly-anionic HA, neutral methyl cellulose and
poly-cationic chitosan undergo no transitions at all. The reason for
these conformational transitions is closely linked to the total
number of axial or equatorial linkages existing in each pyranose
ring (monosaccharide). It is evident that the glycosidic linkages
may act as levers to generate a sufficient torque to undertake the
work done, which is necessary to perform ring conformational
transitions. Hence only glycosidic linkages with axial linkages can
generate enough torque to flip the levers beyond a given critical
stress while equatorial linkages can not. For a more extensive
treatment of the conformational transitions see Marszalek et al.
\cite{Marszalek} for details. This phenomena provides a crucial step
for a molecule to transform from the entropic region into the
Hookean regime. Although the above interpretation of the kinking
conformational changes is largely accepted by most researchers,
there exists some other interpretations to explain these
conformational transitions \cite{Naidoo}.

This paper is divided into four sections. In section~\ref{sec:2}, we
derive a mathematical model for CTs, while in section~\ref{sec:3},
numerical results and some extension work on CTs are developed. In
the last section, we present some conclusions.

\section{Theory}
\label{sec:2}

In this section, we consider some simple applied mathematical models
for a collagen pair. Since CTs contain repeated units of collagen
pairs, a collagen pair is denoted by a single unit of such repeated
collagen pairs, which is illustrated in Figure~\ref{fig:Figure1}.
While we model collagens utilizing Hooke's law, GAGs are modeled
utilizing the worm-like chain model, which has carefully taken the
entropic nature of the molecular chain into account. Due to the
symmetry of CTs, we consider a pair of collagens with axial and
regular GAGs in between, especially bone. In reality, to assemble
such segments into a whole CTs is a challenging task owing to the
complicated molecular interactions between fibrils. In addition, to
simplify our study of external applied forces, which consist of both
tensile and compressive forces, only the tensile stress is examined
because GAGs can convert compression stress into tensile stress and
hence the total tensile stress is assumed to be the vectorial sum of
both tensile and compression forces.

A well constructed and stable collagen pair is assumed to maintain
its structure by the attachment of GAGs, subject to at least a small
perturbation due to molecular interactions between
collagen-collagen, collagens-GAGs, thermal fluctuations, sudden
shocks etc. Suppose that we apply a tensile stress on one end of the
collagen $1$ (see Figure~\ref{fig:Figure2}), where this perturbation
alters the mechanical structure of the collagen pair.

\begin{figure*}
\centering \resizebox{0.45\textwidth}{!}{
\includegraphics{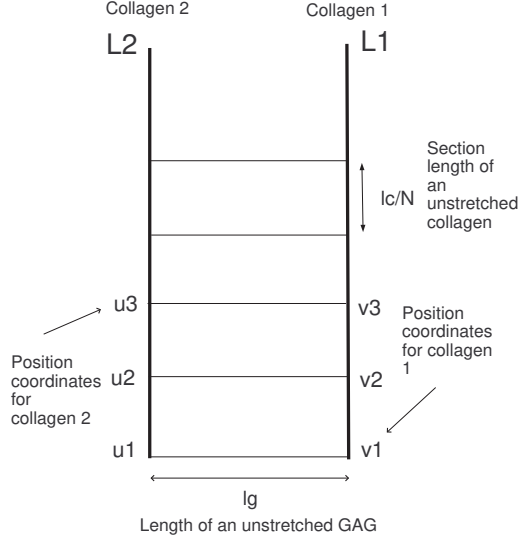}}
\caption{Collagen pair before stretching, where $L_{1}$ and $L_{2}$
denote the collagens 1 and 2 respectively, $v_{1}\ldots v_{N}$ and
$u_{1}\ldots u_{N}$ denote the position coordinates of oligomers in
$L_{1}$ and $L_{2}$ respectively, $lc$ and $lg$ are the natural
lengths of the collagens and GAGs respectively and $N$ is the total
number of oligomers that could exist in each collagen.}
\label{fig:Figure1}
\end{figure*}

\begin{figure*}
\centering \resizebox{0.4\textwidth}{!}{
\includegraphics{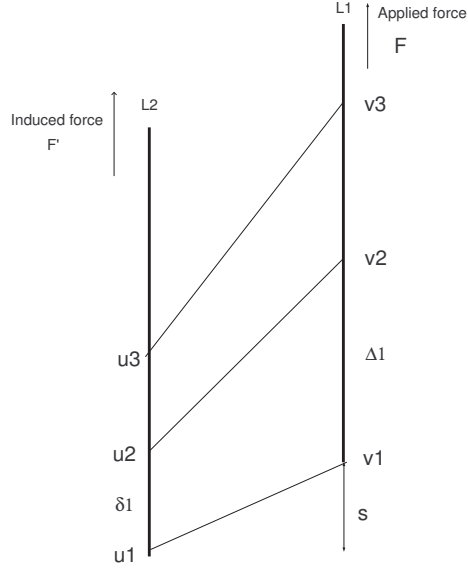}}
\caption{Collagen pair after stretching, by an applied force $F$
assumed to be acting on $L_{1}$ causing an induced force $F'$ in
$L_{2}$, while $s$ denotes the offset length between $L_{1}$ and
$L_{2}$, $\Delta_{1}$ and $\delta_{1}$ denote the length of the
first oligomer in $L_{1}$ and $L_{2}$ respectively and so on.}
\label{fig:Figure2}
\end{figure*}

The displacements between $v_{i+1}$ and $v_{i}$ are denoted by
$\Delta_{i}$ for $i=1, \dots, N-1$ where $N$ is the total number of
oligomers, which are defined by a segment of the collagen pair with
two GAGs attached at its ends. Likewise for the collagen $2$, the
displacements of $u_{i+1}$ and $u_{i}$ are denoted by $\delta_{i}$.
Schematic diagrams of an un-stretched and a stretched collagen pair
are shown in Figure~\ref{fig:Figure1} and Figure~\ref{fig:Figure2}
respectively. We postulate the potential energy of collagens
utilizing Hooke's law, which reads

\begin{displaymath}
V=\frac{1}{2}k\delta ^{2},
\end{displaymath}
where $k$ is a spring constant and $\delta$ is the extension. Given
the potential energy form, it is straightforward to show that the
potential energy of a collagen pair, $V_{c}$, is given by

\begin{equation}
V_{c}=\sum_{i=1}^{N}\frac{1}{2}k_{c}\left(\delta_{i}-\frac{\ell_{c}}{N}\right)^{2}+\sum_{i=1}^{N}\frac{1}{2}k_{c}\left(\Delta_{i}-\frac{\ell_{c}}{N}\right)^{2},\label{1}
\end{equation}
where $N$, $k_{c}$ and $\ell_{c}$ are the total number of oligomers
that exist in the collagen, the spring constant of the collagen, and
the natural length of a collagen respectively. Moreover,
$\delta_{i}$ and $\Delta_{i}$ are defined by
$\delta_{i}=u_{i+1}-u_{i}$ and $\Delta_{i}=v_{i+1}-v_{i}$ for
$i=1,2,\ldots, N-1$ respectively. In addition, we postulate the
potential energy of GAGs utilizing the worm-like chain model. It is
well-known that the interpolation force-extension formula for the
worm-like chain model \cite{Marko} is given by,

\begin{equation}
f=\frac{k_{B}T}{A}\left\{\frac{z}{L}+\frac{1}{4}\left( 1-\frac{z}{L}
\right)^{-2}-\frac{1}{4}\right\}, \label{2}
\end{equation}
where $f$, $k_{B}$, $T$, $A$, $z$, $L$ denote the applied force,
Boltzmann's constant, the absolute temperature, the persistence
length, extension and the total contour length of the GAG
respectively. Then, the potential energy of all GAGs, $V_{g}$, can
then be obtained by integrating Eq.~\ref{2} with respect to $z$ and
summing up the total number of GAGs, $N$, to yield

\begin{eqnarray}
&& V_{g}=
\sum_{i=1}^{N}\bigg\{\frac{1}{2}k_{g}\left\{\sqrt{(v_{i}-u_{i})^{2}+(\ell_{g})^{2}}-\ell_{g}\right\}^{2}
\nonumber\\
&& +\frac{L^{2}}{4}k_{g}\left\{
1-\frac{\sqrt{(v_{i}-u_{i})^{2}+(\ell_{g})^{2}}-\ell_{g}}{L}\right\}^{-1}\nonumber\\
&&- \frac{L}{4}k_{g}\left\{
\sqrt{(v_{i}-u_{i})^{2}+(\ell_{g})^{2}}-\ell_{g}\right\} \bigg\}
,\label{3}
\end{eqnarray}
where $k_{g}$ and $\ell_{g}$ denote $(k_{B}T)/(AL)$ and the natural
length of GAGs respectively. We can then relate $u_{i}$ and $v_{i}$
to $\delta_{i}$, $\Delta_{i}$ and $s$, which is the off-set between
fibrils (See Figure~\ref{fig:Figure2}). Notice that $s$ is a
function of $f$ as the collagen pair starts to slide away with
respect to each other, subject to the external force. Hence, $u_{i}$
and $v_{i}$ are geometrically related by

\begin{eqnarray}
&& v_{1}=u_{1}+s, \nonumber\\
&&
v_{2}=u_{2}+s+(\Delta_{1}-\ell_{c}/N)-(\delta_{1}-\ell_{c}/N)=u_{2}+s+(\Delta_{1}-\delta_{1}),
\nonumber\\
&& \quad\quad \vdots \nonumber\\
&& v_{i}=u_{i}+s+\sum_{k=1}^{i}(\Delta_{k}-\delta_{k}), \nonumber\\
&& \quad\quad \vdots \nonumber\\
&& v_{N}=u_{N}+s+\sum_{k=1}^{N}(\Delta_{k}-\delta_{k}). \label{4}
\end{eqnarray}
After we relate the kinematics between the collagen pair, we can
simplify Eq.~\ref{3} in terms of the displacements $\delta_{i}$ and
$\Delta_{i}$, giving

\begin{eqnarray}
&&V_{g}=\sum_{i=1}^{N}\bigg\{\frac{1}{2}k_{g}\left\{\sqrt{\left[
s+\sum_{k=1}^{i}(\Delta_{k}-\delta_{k})
\right]^{2}+(\ell_{g})^{2}}-\ell_{g}\right\}^{2} \nonumber\\
&&+\frac{L^{2}}{4}k_{g}\left\{ 1-\frac{\sqrt{\left[
s+\sum_{k=1}^{i}(\Delta_{k}-\delta_{k})
\right]^{2}+(\ell_{g})^{2}}-\ell_{g}}{L}\right\}^{-1}\nonumber\\
&&- \frac{L}{4}k_{g}\left\{ \sqrt{\left[
s+\sum_{k=1}^{i}(\Delta_{k}-\delta_{k})
\right]^{2}+(\ell_{g})^{2}}-\ell_{g}\right\}\bigg\}. \label{5}
\end{eqnarray}
Since the elongation of collagen $1$ causes the elongation of
collagen $2$, we can relate $\delta_{i}$ and $\Delta_{i}$ by the
relative displacements $\epsilon_{i}$. That is

\begin{equation}
\Delta_{i}=\delta_{i}+\epsilon_{i}.\label{6}
\end{equation}
Assuming statistical equilibrium, we let $\delta_{i}=\delta$,
$\Delta_{i}=\Delta$ and $\epsilon_{i}=\epsilon$ for all $i$. Hence,
Eqs.~\ref{1} and \ref{5} reduce to

\begin{eqnarray}
&&
V_{c}=\frac{1}{2}k_{c}N\left\{\left(\delta-\frac{\ell_{c}}{N}\right)^{2}+\left(\delta+\epsilon-\frac{\ell_{c}}{N}\right)^{2}\right\},\nonumber\\
&& V_{g}=\sum_{i=1}^{N}\bigg\{\frac{1}{2}k_{g}\left\{
\sqrt{(s+i\epsilon)^{2}+(\ell_{g}})^{2}-\ell_{g}\right\}^{2}\nonumber\\
&& \qquad +\frac{L^{2}}{4}k_{g}\left\{
1-\frac{\sqrt{(s+i\epsilon)^{2}+(\ell_{g}})^{2}-\ell_{g}}{L}\right\}^{-1}\nonumber\\
&& \qquad
-\frac{L}{4}k_{g}\left\{\sqrt{(s+i\epsilon)^{2}+(\ell_{g}})^{2}-\ell_{g}
\right\}\bigg\}. \label{7}
\end{eqnarray}
Since each GAG has its own maximum length $K$ and the displacement
between $u_{N}$ and $v_{N}$ corresponds the maximum length of GAG
given in the collagen pair, which must be smaller or equal to $K$.
Given that, we have

\begin{equation}
|v_{N}-u_{N}|=\sqrt{(s+N\epsilon)^{2}+(\ell_{g})^{2}}\leq K,
\label{8.0}
\end{equation}
from which, we show that the maximum number of oligomers $N_{max}$
allowed in a given collagen satisfies the following inequality

\begin{equation}
N_{max}\leq \frac{\sqrt{(K)^{2}-(\ell_{g})^{2}}-s}{\epsilon} =
\lfloor \frac{\sqrt{(K)^{2}-(\ell_{g})^{2}}-s}{\epsilon}
\rfloor,\label{8}
\end{equation}
where $\lfloor\quad\rfloor$ denotes minimum integer value of the
enclosed real number, e.g. $\lfloor 4.3 \rfloor=4$ respectively.
Note that Eq.~\ref{8} ignores the maximum strain that can be held by
a collagen, and while $\epsilon$ is a rather abstract quantity, it
can be expressed in terms of $s$ by minimizing the maximum potential
energy $E_{max}$ of the collagen pair with respect to $\epsilon$ at
equilibrium (see Appendix for details). At first glance, $s$ is
insignificant under small natural external forces, i.e. thermal
fluctuations and internal molecular interactions. However, $s$
becomes crucial when we consider the collagen pair under large
external tractions. One interesting thing about the above equation
is that it limits the possible number of oligomers that can exist
for a given stable collagen and it is model independent because it
arises solely from a geometric point of view. Also, the inequality
reveals the importance of GAGs in relation to the growth of the
collagen pair. Hence, the longer the maximum length of GAGs, the
longer the structural collagen pair can be. Although the inequality
does not prove the possibility of the existence of ring
conformational transitions of GAGs, it does provide evidence that
the existence of conformational transitions increases $K$ and hence
encourages the growth of the collagen pair. Further, the higher the
effectiveness of the force transmission of GAGs between fibril and
fibril, the lower the value of $\epsilon$, and hence longer the
collagen can be.

The induced force, $F'$, in collagen $2$, generated by collagen $1$
and the relative displacement, $\epsilon$, of the system, can be
determined to be

\begin{equation}
F'=F-k_{c}\epsilon, \quad\quad\quad
\epsilon=\frac{1}{N}(L_{1}-L_{2}), \label{9}
\end{equation}
where $L_{1}$ and $L_{2}$ are the total molecular contour length of
the collagen 1 and 2 respectively. $F'$ and $\epsilon$ can hence be
determined experimentally by knowing $k_{c}$, $F$, $N$, $L_{1}$ and
$L_{2}$. In particular, $\epsilon$ can be measured to find
$N_{max}$. Given $N_{max}$, we can obtain the maximum molecular
energy for a collagen pair, $E_{max}$, namely

\begin{eqnarray}
&&
E_{max}=\frac{1}{2}k_{c}N_{max}\bigg\{\left(\delta-\frac{\ell_{c}}{N_{max}}\right)^{2}+\left(\delta+\epsilon-\frac{\ell_{c}}{N_{max}}\right)^{2}\bigg\}
\nonumber\\ && \qquad
+\sum_{i=1}^{N_{max}}\bigg\{\frac{1}{2}k_{g}\left\{
\sqrt{(s+i\epsilon)^{2}+(\ell_{g}})^{2}-\ell_{g}\right\}^{2}\nonumber\\
&& \qquad +\frac{L^{2}}{4}k_{g}\left\{
1-\frac{\sqrt{(s+i\epsilon)^{2}+(\ell_{g}})^{2}-\ell_{g}}{L}\right\}^{-1}\nonumber\\
&& \qquad
-\frac{L}{4}k_{g}\left\{\sqrt{(s+i\epsilon)^{2}+(\ell_{g}})^{2}-\ell_{g}
\right\}\bigg\}.\label{10}
\end{eqnarray}
The maximum molecular energy $E_{max}$ grows quadratically with the
extension $\delta$ and is related linearly to $N_{max}$, which is
given by Eq.~\ref{8} and determined by $K$ and $\epsilon$. Hence,
the existence of flexible GAGs with long maximum length $K$
increases the number of oligomers $N_{max}$, which in return
tightens the system dramatically. In addition, assuming a large
$N_{max}$ and a small $\epsilon$, the modulus of a collagen pair,
$k_{t}$, can be determined easily by the second derivative of
$E_{max}$ with respect to $\delta$. Upon performing two
differentiations, we approximate the modulus of the collagen pair,
$k_{t}$, as

\begin{equation}
k_{t} \approx k_{c}N_{max}.\label{10.1}
\end{equation}
Again, $k_{t}$ depends linearly on $N_{max}$, where the rest of
arguments are very similar to the discussion for the maximum
molecular energy above.

\section{Numerical results and analysis}
\label{sec:3}
In this section, we carry out a numerical analysis on the results
derived in the previous section, and finally we examine the breaking
point of a collagen pair. Firstly, given Eq.~\ref{10}, we assume the
value of the parameters given in \cite{Redaelli}, namely $k_{c}=0.2$
GPa, $k_{g}=0.02$ GPa, $\ell_{c}=100$ $\mu$m, $\ell_{g}=0.02$ nm,
$L=0.2$ nm and $3$ maximum numbers of oligomer are examined, namely
$N_{max1}=1$ K, $N_{max2}=10$ K and $N_{max3}=0.1$ M. Further,
without loss of generality, the offset $s$ is assumed to be zero,
which corresponds to the tight molecular interaction between fibrils
or the collagen pair under a small traction. The potential energy of
GAGs is neglected due to its insignificance in comparison to the
potential energy of the collagen pair that is demonstrated in
Figure~\ref{fig:Figure3}, where $\epsilon$ has been linearized with
the extension $\delta$. The numerical result shows that the
potential energy of GAGs utilizing the above parameters is $6$
orders of magnitude smaller than the potential energies of the
paired collagen for all three cases we considered. Given that, the
potential energies of the collagen pair versus the extension
$\delta$, ranging from $0$ to $100$ nm for $N_{max1}$, $N_{max2}$
and $N_{max3}$, are plotted together in Figure~\ref{fig:Figure4}.

\begin{figure*}
\centering \resizebox{0.75\textwidth}{!}{
\includegraphics{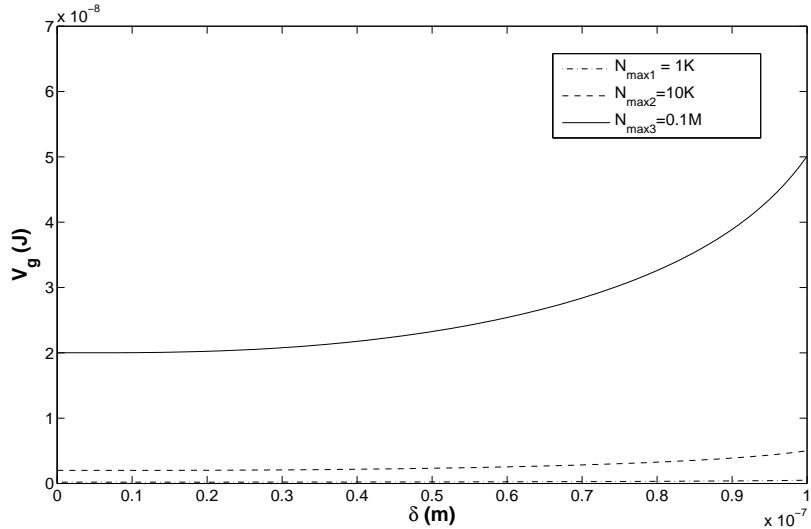}}
\caption{Potential energy of GAG, $V_{g}$, versus extension,
$\delta$ ranging from $0$ to $100$ nm, for $N_{max1}=1$ K,
$N_{max2}=10$ K and $N_{max3}=0.1$ M respectively.}
\label{fig:Figure3}
\end{figure*}

\begin{figure*}
\centering \resizebox{0.75\textwidth}{!}{
\includegraphics{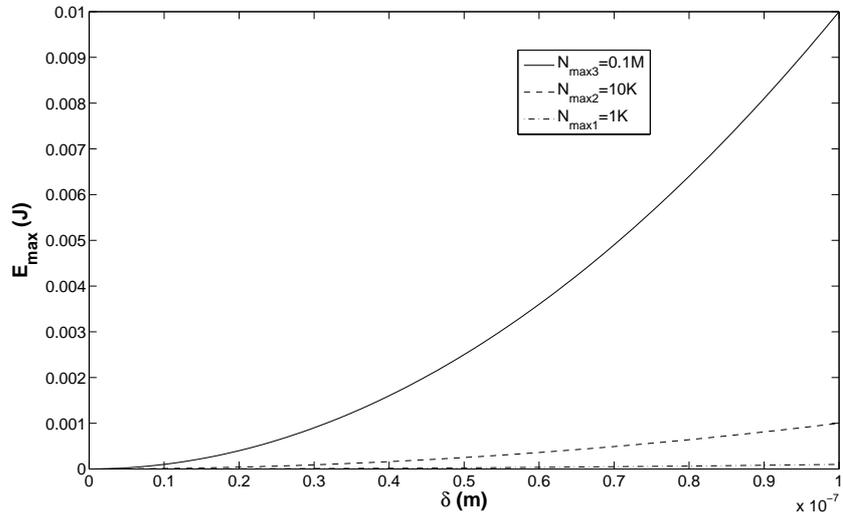}}
\caption{Potential energy of the collagen pair, $E_{max}$, versus
extension, $\delta$ ranging from $0$ to $100$ nm, for $N_{max1}=1$
K, $N_{max2}=10$ K and $N_{max3}=0.1$ M respectively.}
\label{fig:Figure4}
\end{figure*}

\begin{figure*}
\centering \resizebox{0.6\textwidth}{!}{
\includegraphics{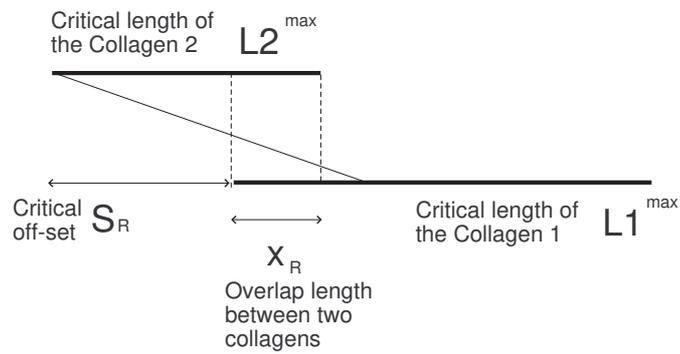}}
\caption{The breaking fraction of the collagen pair}
\label{fig:Figure5}
\end{figure*}

Note that the maximum molecular energy increases quadratically when
the extension increases linearly. In addition, $\delta$ is
determined by the applied force $F$ and the relative displacement
$\varepsilon$. For example, for a constant $\Delta$, $\delta$
achieves its maximum value when $\varepsilon=0$. Further, $E_{max}$
increases linearly with $N_{max}$, which in return depends
positively on $K$ but negatively on $\varepsilon$. The toughness of
a collagen pair, $k_{t}$, is defined in the previous section as the
second derivative of $E_{max}$ with respect to $\delta$, and hence
the higher the slope of $E_{max}$ with respect to the extension
$\delta$, the stiffer the collagen pair is. In conclusion, given a
collagen pair, it is tougher whenever GAGs have a higher value of
$K$ and a higher ability in transmitting forces between fibrils.
However, the existence of GAGs contributes almost nothing to the
toughness of the collagen pair but is significant to maintain the
stable structure and boost up the toughness of the collagen pair,
which are consistent to the results that an increased PYD or DPD
ratio (the most abundant mature GAGs in bone collagen) is related to
the increased compressive strength in bone (CTs)
\cite{Lees,Bailey,Oxlund1,Oxlund2,Banse} but has no huge effect on
toughness or ductility of bone
\cite{Zioupos,Wang,Keaveny,Hernandez}.

We end this section by extending concepts that we developed above to
determine the point at which a collagen pair will break. For the
study of a collagen pair, an important parameter is the fraction of
the collagen pair, $\chi_{R}$, in which the breaking point occurs
(see Figure~\ref{fig:Figure5}). The breaking point is assumed to
occur when the bridge between $u_{1}$ and $v_{1}$ is finally broken
(see Figures~\ref{fig:Figure1} and ~\ref{fig:Figure2}). From
Figure~\ref{fig:Figure5}, we write

\begin{equation}
L_{1}^{max}+L_{2}^{max}-\chi_{R}=s_{R}+L_{1}^{max},\label{11}
\end{equation}
where $L_{1}^{max}$ and $L_{2}^{max}$ are the critical lengths of
collagens 1 and 2 respectively and $s_{R}$ is the critical off-set
of the system. After some re-arrangement, we find that

\begin{equation}
\chi_{R}=L_{2}^{max}-s_{R}.\label{12}
\end{equation}
In addition, we know that when the collagen pair is about to be torn
apart, $N_{max}=0$, from Eq.~\ref{8}, we deduce

\begin{equation}
s_{R}=\sqrt{(K)^{2}-(\ell_{g})^{2}}. \label{13}
\end{equation}
Therefore, upon comparing Eqs.~\ref{12} and \ref{13}, we have

\begin{equation}
\chi_{R}=L_{2}^{max}- \sqrt{(K)^{2}-(\ell_{g})^{2}}. \label{14}
\end{equation}
Notice that the above equation links the macro-quantity $\chi_{R}$
to micro-quantities such as $L_{2}^{max}$, $K$ and $\ell_{g}$. Once
again, the larger the possible length of GAGs, the smaller the
breaking fraction of the collagen pair, which implies that we need
to put more forces in the system to tear the collagen pair apart,
which is consistent to the theoretical result obtained by
\cite{Buehler}.

\section{Conclusion}
\label{sec:4} In this paper, we utilize simple applied mathematical
modeling techniques, for example Hooke's law, to describe the
structure of a collagen pair. We find that for maintaining the
stability of a collagen pair, the maximum number of oligomers that
can exist in a given collagen depends on the maximum length of GAGs
and the effectiveness of the GAGs in transferring forces between
fibrils. That is, a collagen can grow longer for the longer GAGs and
the higher effectiveness of GAGs in transferring forces between
fibrils. This concept can then be extended naturally to the
toughness and the breaking point of a collagen pair, which are found
to be intimately related to the characteristics of GAGs. In
addition, the possible conformational transitions of GAGs strengthen
the whole structure of CTs and as such CTs maintain our body shape
from any large external traction. We note that our work does not
give the complete picture where a more sophisticated theoretical
investigation of the interactions between collagens and GAGs needs
to be undertaken and extend our model to investigate the mechanical
properties of CTs. In addition, we sincerely refer our readers, who
are interested in utilizing computing simulations and continuum
mechanics to model the macroscopic structure of CTs, to those
references listed in the introduction.

\section*{Appendix}
From Eq.~\ref{10}, if $N_{max}$ is sufficiently large, then we can
approximate the summation by integration. Upon utilizing equation
Eq.~\ref{8}, we obtain

\begin{eqnarray}
&&
E_{max}=\frac{1}{2}k_{c}\left(\frac{\Phi-s}{\epsilon}\right)\bigg\{\left(\delta-\frac{\ell_{c}}{\Phi-s}\epsilon\right)^{2}+
\left[\delta+\left(1-\frac{\ell_{c}}{\Phi-s}\right)\epsilon\right]^{2}\bigg\}
\nonumber\\
&& +
\int_{0}^{\frac{\Phi-s}{\epsilon}}\frac{1}{2}k_{g}\left\{\sqrt{(s+\epsilon
x)^{2}+\ell_{g}^{2}}-\ell_{g}\right\}^{2}dx \nonumber\\
&&+ \int_{0}^{\frac{\Phi-s}{\epsilon}} \frac{L^{2}}{4} k_{g} \left\{
1-\frac{\sqrt{(s+\epsilon
x)^{2}+\ell_{g}^{2}}-\ell_{g}}{L}\right\}^{-1}dx\nonumber\\
&&-\int_{0}^{\frac{\Phi-s}{\epsilon}}\frac{L}{4}k_{g}\left\{\sqrt{(s+\epsilon
x)^{2}+\ell_{g}^{2}}-\ell_{g}\right\}dx, \label{A1}
\end{eqnarray}
where $\Phi=\sqrt{K^{2}-\ell_{g}^{2}}$. If we substitute $s+\epsilon
x=\ell_{g}\tan\theta$ and perform the integration, we obtain

\begin{equation}
E_{max}=\frac{1}{2}k_{c}\left(\frac{\Phi-s}{\epsilon}\right)
\bigg\{\left(\delta-\frac{\ell_{c}}{\Phi-s}\epsilon\right)^{2}+
\left[\delta+\left(1-\frac{\ell_{c}}{\Phi-s}\right)\epsilon\right]^{2}\bigg\}+\frac{H(s)}{\epsilon},\label{A2}
\end{equation}
where
$H(s)=(1/2)k_{g}\ell_{g}^{3}a(s)+(L^{2}/4)k_{g}\ell_{g}b(s)-(L/4)k_{g}\ell_{g}^{2}c(s)$
and

\begin{eqnarray}
&& a(s)=2\tan\theta + (1/3)\tan^{3}\theta-\tan\theta\sec\theta
\nonumber\\
&&- \ln
(\sec\theta+\tan\theta)|_{\arctan(s/\ell_{g})}^{\arctan(\Phi/\ell_{g})},
\nonumber\\
&& b(s)= -\frac{L\ln\left[\tan\frac{\theta}{2}+1
\right]}{\ell_{g}}+\frac{L\ln\left[\tan\frac{\theta}{2}-1
\right]}{\ell_{g}} \nonumber\\
&& +2L\left(1+\frac{L}{\ell_{g}}\right)\frac{\arctan\left[
\frac{(L+2\ell_{g})\tan\frac{\theta}{2}}{\sqrt{L(L+2\ell_{g})}}
\right]}{\sqrt{L(L+2\ell_{g})}}|_{\arctan(s/\ell_{g})}^{\arctan(\Phi/\ell_{g})}, \nonumber\\
&& c(s)=
\frac{1}{2}\tan\theta+\frac{1}{2}\ln(\tan\theta+\sec\theta)-\tan\theta|_{\arctan(s/\ell_{g})}^{\arctan(\Phi/\ell_{g})}.\nonumber\\
&& \label{A3}
\end{eqnarray}
We note that the energy form given in Eq.~\ref{A2} might be utilized
to carry out computer simulations or mathematical modelings for CTs
in a more accurate way. To find the minimum value of $E_{max}$ with
respect to $\epsilon$, we require $\partial
E_{max}/\partial\epsilon=0$. Given that, we have

\begin{equation}
\epsilon = \sqrt{\frac{2\left[ \delta^{2} +
\frac{H}{k_{c}(\Phi-s)}\right]}{\left[ \left(\frac{\ell_{c}}{\Phi-s}
\right)^{2}+\left(1-\frac{\ell_{c}}{\Phi-s}\right)^{2}
\right]}}.\label{A4}
\end{equation}
Upon knowing $s$, $\epsilon$ can be easily solved from the above
equation.

\section*{Acknowledgements}
We gratefully acknowledge the support from the Discovery Project
Scheme of the Australian Research Council.


\begin{thebibliography}{99}
\bibitem {Scott1991} Scott J E (1991) Proteoglycan: collagen interactions and cor-neal ultrastructure. Biochem.Soc.Trans. 19: 877-881
\bibitem {Scott1988} Scott J E (1988) Proteoglycan-fibrillar collagen interactions. Biochem.J. 252: 313-323
\bibitem {Svoboda} Svoboda K, Schmidt C F, Schnapp B J, Block S M (1993) Direct Observation of kinesin stepping by optical trapping interferometry. Nature 365:721-727
\bibitem {Haverkamp2005} Haverkamp R G, Williams M A K, Scott J E (2005) Stretching single molecules of connective tissue glycans to characterize their shape-maintaining elasticity. Biomacromolecules 6: 1816-1818
\bibitem {Scott} Scott J E (1975) Composition and structure of the pericellular environment: physiological function and chemical composition of pericellular proteoglycan (an evolutionary view). Philos. Trans. R. Soc. Lond. B. Biol. Sci. 271: 235-242
\bibitem {Burr} Burr D B (2002) The contribution of the organic matrix to bone's materials properties. Bone 31: 8-11
\bibitem {Redaelli} Redaelli A, Vesentini S, Soncini M, Vena P, Mantero S, Montevecchi F M (2003) Possible role of decorin glycosaminoglycans in fibril to fibril force transfer in relative mature tendons-a computational study from molecular to microstructural level. J. Biomech. 36: 1555-1569
\bibitem {Akkus} Akkus O (2005) Elastic deformation of mineralized collagen fibrils: an equivalent inclusion based composite model.  J.Biomech.Eng. 127: 383-390
\bibitem {Fritsch} Fritsch A, Hellmich C (2007) `Universal' microstructural patterns in cortical and trabecular, extracellular bone materials, micromechanics-based prediction of anisotropic elasticity.  J. Theor. Biol. 244: 597-620
\bibitem {Hellmich} Hellmich C, Barthelemy J F, Dormieux L (2004) Mineral-collagen interactions in elasticity of bone ultrastructure-a continuum micromechanics approach.  Eur. J. Mech. A. Solids 23: 783-810
\bibitem {Jager} Jager I, Fratzl P (2000) Mineralized collagen fibrils: a mechanical model with a staggered arrangement of mineral particle. Biophys. J. 79: 1737-1746
\bibitem {Ji} Ji B, Gao H (2004) Mechanical properties of nanostructure of biological materials. J. Mech. Phys. Solids 52: 1963-1990
\bibitem {Kotha} Kotha S P, Guzelsu N (2003) Effect of bone mineral content on the tensile properties of cortical bone: experiments and theory. J. Biomech. Eng. 125: 785-793
\bibitem {Wang1} Wang X, Qian C (2006) Prediction of microdamage formation using a mineral-collagen composite model. J. Biomech. 39: 595-602
\bibitem {Baumann} Baumann C G, Bloomfield V A, Smith S B, Bustamante C, Wang M D, Block S M (2000) Stretching of single collapsed DNA molecules. Biophys. J. 78: 1965-1978
\bibitem {Rief1} Rief M, Oesterhelt F, Heymann B, Gaub H E (1997) Single molecule force spectroscopy on polysaccharides by atomic force microscopy. Science 275: 1295-1297
\bibitem {Rief2} Rief M, Fernandez J M, Gaub H E (1998) Elastically coupled two-level systems as a model for biopolymer extensibility. Phys. Rev. Lett. 81: 21
\bibitem {Oesterhelt} Oesterhelt F, Rief M, Gaub H E (1999) Single molecule force spectroscopy by AFM indicates helical structure of poly(ethylene-glycol) in water. New J. Phys. 1: 6.1-6.11
\bibitem {Barrows} Barrows S E, Dulles F J, Cramer C J, French A D, Truhlar D G (1995) Relative stability of alternative chair forms and hydroxymethyl conformations of $\beta$-$_{D}$-glucopyranose. Carbohydr. Res. 276: 219-251
\bibitem {Joshi} Joshi N V, Rao V S R (1979) Flexibility of the pyranose ring in $\alpha$ and $\beta$ -glucoses. Biopolymers 18: 2993-3004
\bibitem {Dowd} Dowd M K, French A D, Reilly P J (1994) Modeling of aldopyranosyl ring puckering with MM3. Carbohydr. Res. 264: 1-19
\bibitem {Marszalek} Marszalek P E, Pang Y P, Li H, Yazal J E, Oberhauser A F, Fernandez J M (1999) Atomic levers control pyranose ring conformations. PNAS 96: 7894-7898
\bibitem {Naidoo} Kuttel M, Naidoo K J (2005) Glycosidic linkage rotations determine amylose stretching mechanism. J. Am. Chem. Soc. 127: 12-13
\bibitem {Marko} Marko J F, Siggia E D (1995) Stretching DNA. Macromolecules 28: 8759
\bibitem {Lees} Lees S, Eyre D R, Barnard S M (1990) BAPN dose dependence of mature crosslinking in bone marix collagen of rabbit compact bone: corresponding variation of sonic velocity and equatorial diffraction spacing. Connective Tissue Research 24: 95-105
\bibitem {Bailey} Bailey A J, Wotton S F, Sims T J, Thompson P W (1992) Post-translational modifications in the collagen of human osteoporotic femoral head. Biochem. Biophys. Res. Commun. 185: 801-805
\bibitem {Oxlund1} Oxlund H, Barckmann M, Ortoft G, Ancreassen T T (1995) Reduced concentrations of collagen cross-links are associated with reduced strength of bone. Bone 17: 365S-371S
\bibitem {Oxlund2} Oxlund H, Mosekilde Li, Ortoft G (1996) Reduced concentration of collagen reducible cross links in human trabecular bone with respect to age and osteoporosis. Bone 19: 479-484
\bibitem {Banse} Banse X, Sims T J, Bailey A J (2002) Mechanical properties of adult vertebral cancellous bone: correlation with collagen intermolecular corss-links. J. Bone Miner. Res. 17: 1621-1628
\bibitem {Zioupos} Zioupos P, Currey J D, Hamer A J (1999) The role of collagen in the declining mechanical properties of aging human cortical bone. J. Biomed. Mater. Res. 45: 108-116
\bibitem {Wang} Wang X, Shen X, Li X, Agarwal C M (2002) Age-related changes in the collagen network and toughness of bone. Bone 31: 1-7
\bibitem {Keaveny} Keaveny T M, Morris G E, Wong E K, Yu M, Sakkee A N, Verzijl N, Bank R A (2003) Collagen status and brittleness of human cortical bone in the elderly. J. Bone Miner. Res. 18: S307
\bibitem {Hernandez} Hernandez C J, Tang S Y, Baumbach B M, Hwu P B, Sakkee A N, van der Ham F, DeGroot J, Bank R A, Keaveny T M (2005) Trabecular microfracture and the influence of pyridinium and non-enzymatic glycation-mediated collagen cross-links. Bone 37: 825-832
\bibitem {Buehler} Buehler M J (2006) Nature designs tough collagen: Explaining the nanostructure of collagen fibrils. PNAS 103: 12285-12290
\end{thebibliography}
\end{document}